\documentclass[nofootinbib,twocolumn,showpacs,superscriptaddress,preprintnumbers,prd]{revtex4-1}
\usepackage{graphicx,amsmath,amssymb,dcolumn,bm}
\usepackage{epic,eepic}
\usepackage[dvipdfmx]{pict2e}
\begin{document}
\preprint{KEK-TH-1912, J-PARC-TH-0054}
\title{Tensor-polarized structure function {\boldmath$b_1$} \\
       in the standard convolution description of the deuteron}
\author{W. Cosyn}
\affiliation{Department of Physics and Astronomy, Ghent University, 
             Proeftuinstraat 86, B9000 Ghent, Belgium}
\author{Yu-Bing Dong}
\affiliation{Institute of High Energy Physics, 
             Chinese Academy of Sciences, Beijing 100049, China}
\affiliation{Theoretical Physics Center for Science Facilities (TPCSF), 
             CAS, Beijing 100049, China}
\author{S. Kumano}
\affiliation{KEK Theory Center,
             Institute of Particle and Nuclear Studies, \\
             High Energy Accelerator Research Organization (KEK), \\
             1-1, Ooho, Tsukuba, Ibaraki, 305-0801, Japan}
\affiliation{J-PARC Branch, KEK Theory Center,
             Institute of Particle and Nuclear Studies, KEK, \\
           and
           Theory Group, Particle and Nuclear Physics Division, 
           J-PARC Center, \\
           203-1, Shirakata, Tokai, Ibaraki, 319-1106, Japan}
\author{M. Sargsian}
\affiliation{Department of Physics, Florida International University, 
             Miami, Florida 33199, USA} 
\date{March 31, 2017}
\begin{abstract}
Tensor-polarized structure functions of a spin-1 hadron are 
additional observables which do not exist for the spin-1/2 nucleon. 
They could probe novel aspects of the internal hadron structure.
Twist-2 tensor-polarized structure functions are $b_1$ and $b_2$,
and they are related by the Callan-Gross-like relation 
in the Bjorken scaling limit. In this work, we theoretically calculate 
$b_1$ in the standard convolution description for the deuteron.
Two different theoretical models, a basic convolution description and 
a virtual nucleon approximation, are used for calculating $b_1$ 
and their results are compared with the HERMES measurement. 
We found large differences between our theoretical results and the data. 
Although there is still room to improve by considering higher-twist effects 
and in the experimental extraction of $b_1$ from the spin asymmetry $A_{zz}$, 
there is a possibility that the large differences require
physics beyond the standard deuteron model for their interpretation.
Future $b_1$ studies could shed light on a new field of hadron physics. 
In particular, detailed experimental studies of $b_1$ will start soon 
at the Thomas Jefferson National Accelerator Facility. 
In addition, there are possibilities to investigate tensor-polarized 
parton distribution functions and $b_1$ at Fermi National Accelerator 
Laboratory and a future electron-ion collider. Therefore, 
further theoretical studies are needed for understanding
the tensor structure of the spin-1 deuteron, including a new mechanism 
to explain the large differences between the current data 
and our theoretical results.
\end{abstract}
\pacs{13.60.Hb, 13.88.+e}
\maketitle

\section{Introduction}
\label{intro}
\vspace{-0.20cm}

Spin structure of the nucleon has been investigated extensively
especially after the European Muon Collaboration discovery on the small 
quark-spin contribution to the nucleon spin. Now, its studies are focused 
on gluon-spin and orbital-angular-momentum effects. On the other hand,
a spin-1 hadron has richer spin structure than the spin-1/2 nucleon
in the sense that there are four additional structure functions
in charged-lepton inclusive deep inelastic scattering (DIS) \cite{fs83,hjm89}.
They are named $b_1$, $b_2$, $b_3$, and $b_4$ \cite{hjm89}, 
which are associated with the tensor structure of the spin-1
hadron. The leading-twist structure functions are $b_1$ and $b_2$, and
they are related to each other by the Callan-Gross-like relation
$2 x_D b_1=b_2$, where $x_D$ is the scaling variable for the spin-1 hadron,
in the Bjorken scaling limit.
These additional structure functions are interesting quantities 
for probing different dynamical aspects of hadron structure, 
possibly of exotic nature as we suggest in this article, 
from the ones for the spin-1/2 nucleon.

Within the parton model, the structure function $b_1$ satisfies
the sum $\int dx \, b_1 (x) =0$ \cite{b1-sum}, where $x$ is
the Bjorken scaling variable, by considering only the valence-quark part
for the tensor structure.
However, it does not mean $b_1 (x) =0$ for actual hadrons. 
In the fixed-target DIS, the simplest stable 
spin-1 target is the deuteron. If the deuteron $b_1$ is calculated
in the convolution model \cite{hjm89,kh91}, it is, in fact, finite
and shows an oscillatory behavior as a function of $x$.
Furthermore, shadowing mechanisms contribute significantly 
to $b_1$ at small $x$ \cite{b1-shadowing,epw-1997}, and pions
in the deuteron could also play a role \cite{miller-b1}.
There are related studies to the spin-1 hadron structure on 
a polarized proton-deuteron Drell-Yan process 
\cite{pd-drell-yan,ks-2016,Fermilab-dy},
leptoproduction of a spin-one hadron \cite{rho-production},
fragmentation functions \cite{spin-1-frag}, 
generalized parton distributions \cite{spin-1-gpd}, 
target-mass corrections \cite{mass-corr},
positivity constraints \cite{dmitrasinovic-96},
lattice QCD estimates \cite{lattice},
and angular momenta for the spin-1 hadron \cite{angular-spin-1}.
The spin-1 deuteron structure can be also investigated by tagging 
the final state proton \cite{tagged-spin-1}.
In addition, it is a unique opportunity to investigate
the gluon transversity distribution which exists only for hadrons 
with spin$\ge 1$ \cite{trans-g}.

The first measurement of $b_1$ was reported by the HERMES 
Collaboration in 2005 \cite{hermes05}, and possible tensor-polarized
parton distribution functions (PDFs) were extracted from the data
\cite{tensor-pdfs}. The HERMES data are much larger in magnitude
compared to the conventional convolution calculation of Refs.~\cite{hjm89,kh91}.
It indicates that a new hadron mechanism should possibly be 
considered to interpret the large magnitude of $b_1$. 
As such an exotic mechanism, a contribution from a hidden-color 
state is proposed as a possibility together with
a pionic contribution in Ref.~\cite{miller-b1}.

The deuteron tensor structure has been investigated 
for a long time at low energies in terms of hadron
degrees of freedom, and it originates from the D-state
admixture in a bound proton-neutron system.
However, time has come to investigate the tensor structure 
in terms of quark and gluon degrees of freedom
through the structure functions $b_{1-4}$.
In particular, the HERMES data seem to suggest
a possible existence of an exotic hadron mechanism 
for interpreting their data because they deviate significantly
from a conventional theoretical prediction. 
For describing nuclear structure functions at medium and large $x$, 
it is standard to use a convolution formalism, where
a nuclear structure function is given by the corresponding
one convoluted with a nucleon momentum distribution
in a nucleus \cite{basic-conv,ek-2003,nuclear-sfs}. 
It is considered as a baseline calculation in describing 
nuclear modifications of $F_2$ at medium and large $x$
in terms of the nuclear binding, Fermi motion, and 
short-range correlations embedded in the spectral function 
of the nucleon. We can use the same model for the deuteron
in describing the structure functions including $b_1$.
In addition, we can also use another convolution description
of the virtual nucleon approximation 
\cite{Frankfurt:1981mk,Keister:1991sb,cms-vna} 
which is used for describing the tagged structure functions of
the spin-1 deuteron \cite{tagged-spin-1}.

There is only one type of theoretical calculation in the convolution 
picture. A basic formalism was shown in Ref. \cite{hjm89}, 
and updated result is provided in Ref. \cite{kh91}.
Since the deviation from this model is very important
for indicating a new hadron-physics mechanism, we need 
to check its results independently. 
This is the purpose of this article. In fact, we obtain
very different numerical results from the previous 
theoretical estimate of Ref. \cite{kh91}
as shown in Sec.\,\ref{results}.
This work is important for considering the upcoming 
Thomas Jefferson National Accelerator Facility (JLab) experiments
\cite{Jlab-b1,azz}. There are also experimental possibilities to investigate
the tensor-polarized PDFs and structure functions
at Fermilab \cite{pd-drell-yan,ks-2016,Fermilab-dy}
and the future electron-ion collider \cite{eic}.

We use two types of convolution formalisms for calculating $b_1$ 
of the deuteron. One is a basic one in describing the nuclear structure
functions as explained, for example, in Refs. \cite{basic-conv,ek-2003,nuclear-sfs}.
The other is the virtual nucleon approximation in 
Refs. \cite{Frankfurt:1981mk,Keister:1991sb,cms-vna,tagged-spin-1}
by considering that the virtual photon interacts with an off shell
nucleon and another spectator nucleon is on mass shell.
Consequently, the obtained structure function $b_1$ from both models is 
compared with
the HERMES measurements.

In this article, we first introduce the tensor structure functions
$b_{1-4}$ in Sec.\,\ref{b1}, and convolution formalism
for $b_1$ is explained in Sec.\,\ref{formalism}.
Numerical results are shown in Sec.\,\ref{results},
and they are summarized in Sec.\,\ref{summary}.

\vspace{+0.5cm}

\section{Tensor-polarized structure function $\bm{b}_{\bm{1}}$}
\label{b1}

\vspace{-0.3cm}

\begin{figure}[h]
\vspace{-0.30cm}
\begin{center}
   \includegraphics[width=4.0cm]{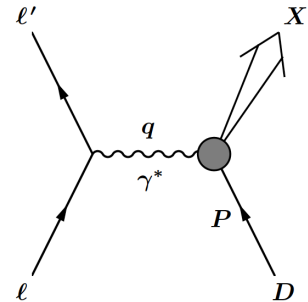}
\end{center}
\vspace{-0.60cm}
\caption{Charged-lepton DIS from spin-1 hadron.}
\label{fig:dis-spin1}
\vspace{-0.5cm}
\end{figure}

We introduce tensor-polarized structure functions for a spin-1 hadron
in charged-lepton DIS as shown in Fig.\,\ref{fig:dis-spin1}.
The initial and final lepton momenta
are $\ell$ and $\ell '$, respectively, 
$q \, (=\ell -\ell ')$ is the momentum transfer, and $P$ is the spin-1
hadron momentum. 
The cross section is described by a lepton tensor multiplied by
the hadron tensor $W_{\mu \nu}$ expressed in terms of
eight structure functions as \cite{hjm89,kk08,sk14}
\begin{widetext}
\vspace{-0.40cm}
\begin{align}
W_{\mu \nu}^{\lambda' \lambda} (P,q)
= &  \frac{1}{4 \pi} 
               \int d^4 \xi \, e^{i q \cdot \xi} \,
               \langle \, P, \lambda' \, | \, [ \, J_\mu^{\, em} (\xi) ,
                      J_\nu^{\, em} (0) ]  \, | \, P, \lambda \, \rangle
\nonumber \\[-0.15cm]
   = & -F_1 \hat{g}_{\mu \nu} 
     +\frac{F_2}{M \nu} \hat{P}_\mu \hat{P}_\nu 
     + \frac{ig_1}{\nu}\epsilon_{\mu \nu \lambda \sigma} q^\lambda S^\sigma  
     +\frac{i g_2}{M \nu ^2}\epsilon_{\mu \nu \lambda \sigma} 
      q^\lambda (P \cdot q \, S^\sigma - S \cdot q \, P^\sigma )
\notag \\[-0.02cm]
& 
     -b_1 r_{\mu \nu} 
     + \frac{1}{6} b_2 (s_{\mu \nu} +t_{\mu \nu} +u_{\mu \nu}) 
     + \frac{1}{2} b_3 (s_{\mu \nu} -u_{\mu \nu}) 
     + \frac{1}{2} b_4 (s_{\mu \nu} -t_{\mu \nu}) ,
\label{eqn:w-1}
\end{align}
\vspace{-0.10cm}
where the new tensor-polarized structure functions are $b_{1-4}$,
which do not exist for the spin-1/2 nucleon. 
The coefficients $r_{\mu \nu}$, $s_{\mu \nu}$, 
$t_{\mu \nu}$, and $u_{\mu \nu}$ are defined 
by the spin-1 polarization vector $E^\mu$, 
hadron and virtual-photon momenta ($P$, $q$),
and initial and final spin states ($\lambda$, $\lambda'$) as
\vspace{-0.10cm}
\begin{align}
r_{\mu \nu} = & \frac{1}{\nu ^2}
   \bigg [ q \cdot E ^* (\lambda') q \cdot E (\lambda) 
           - \frac{1}{3} \nu ^2  \kappa \bigg ]
   \hat{g}_{\mu \nu}, 
\ \ \ \ 
s_{\mu \nu} =  \frac{2}{\nu ^2} 
   \bigg [ q \cdot E ^* (\lambda') q \cdot E (\lambda) 
           - \frac{1}{3} \nu ^2  \kappa \bigg ]
\frac{\hat{P}_\mu \hat{P}_\nu}{M \nu}, \notag \\
t_{\mu \nu} = & \frac{1}{2 \nu ^2}
   \bigg [ q \cdot E ^* (\lambda') 
           \left\{ \hat{P}_\mu \hat E_\nu (\lambda) 
                 + \hat{P} _\nu \hat E_\mu (\lambda) \right\}
   + \left\{ \hat{P}_\mu \hat E_\nu^* (\lambda')  
           + \hat{P}_\nu \hat E_\mu^* (\lambda') \right\}  
     q \cdot E (\lambda) 
   - \frac{4 \nu}{3 M}  \hat{P}_\mu \hat{P}_\nu \bigg ] ,
\notag \\
u_{\mu \nu} = & \frac{M}{\nu} 
   \bigg [ \hat E_\mu^* (\lambda') \hat E_\nu (\lambda) 
          +\hat E_\nu^* (\lambda') \hat E_\mu (\lambda) 
   +\frac{2}{3}  \hat{g}_{\mu \nu}
   -\frac{2}{3 M^2} \hat{P}_\mu \hat{P}_\nu \bigg ] ,
\end{align}
\end{widetext}
where the states $\lambda$ and $\lambda'$ are explicitly denoted
for describing higher-twist contributions by off diagonal terms 
with $\lambda' \ne \lambda$ \cite{hjm89}.
Here, $\hat{g}_{\mu \nu}$ and $\hat{X}_\mu$ ($=\hat{P}_\mu$, $\hat{E}_\mu$)
are defined as
\begin{align}
\hat{g}_{\mu \nu} \equiv  g_{\mu \nu} -\frac{q_\mu q_\nu}{q^2}, \ \ 
\hat{X}_\mu \equiv X_\mu -\frac{X \cdot q}{q^2} q_\mu ,
\label{eqn:hat}
\end{align}
to satisfy the current conservation 
$q^\mu W _{\mu \nu} = q^\nu W _{\mu \nu}=0$,
$M$ is the hadron mass, $\nu$ and $Q^2$ are defined by $\nu ={P \cdot q}/{M}$, 
$Q^2=-q^2 >0$, 
$\epsilon_{\mu \nu \lambda \sigma}$ is an antisymmetric 
tensor with the convention $\epsilon_{0123}=+1$,
$\kappa$ is defined by $\kappa= 1+{Q^2}/{\nu^2}$, and
$S^\mu$ is the spin vector of the spin-1 hadron.
The polarization vector satisfies the constraints, $P \cdot E =0$ and $E^* \cdot E =-1$,
and it is taken as the spherical unit vectors \cite{edmond}
for the spin-1 hadron at rest.
It is related to the spin vector by
\begin{equation}   
(S_{\lambda' \lambda})^{\mu}
      = -\frac{i}{M} \epsilon ^{\mu \nu \alpha \beta} 
                E^*_\nu (\lambda') E_\alpha (\lambda) P_\beta .
\end{equation}

The structure functions $b_1$ and $b_2$ are twist-2 structure functions,
and they are related to each other by the Callan-Gross-like relation 
$2x_D b_1=b_2$ in the Bjorken scaling limit. The functions $b_3$ and $b_4$
are twist-4 ones, so that the leading structure functions 
would be investigated first. In the parton model, $b_1$ is expressed
in terms of the tensor-polarized parton distribution functions
$\delta_{_T} f$ as \cite{delta-T-notation}
\begin{align}
b_1 (x,Q^2) & = \frac{1}{2} \sum_i e_i^2 
      \, \left [ \delta_{_T} q_i (x,Q^2) 
      + \delta_{_T} \bar q_i (x,Q^2)   \right ] , 
\nonumber \\
\! \!
\delta_{_T} f (x,Q^2) & \equiv f^0 (x,Q^2)
          - \frac{f^{+1} (x,Q^2) +f^{-1} (x,Q^2)}{2},
\label{eqn:b1-parton}
\end{align}
where $f^\lambda$ is an unpolarized parton distribution
in the hadron spin state $\lambda$, and
$e_i$ is the charge of the quark flavor $i$.
The Bjorken scaling variable $x$ defined
\begin{align}
x =\frac{Q^2}{2 M_N \nu}  , \ \ \ 
\label{eq:x}
\end{align}
where $M_N$ is the nucleon mass, and
the scaling variable could be defined
as $x_{_D} = Q^2/(2P \cdot q)$ for the deuteron
so as to satisfy the kinematical condition $0<x_{_D}<1$.
For the fixed-target deuteron, they are related to each other by
$x = x_{_D} M / M_N \simeq 2 x_D$, so that the range of 
the Bjorken variable becomes $0<x \lesssim 2$. So far, 
the variable $x$ is used for showing experimental data 
of deuteron structure functions.
At this stage, there is no DIS measurement at a large-enough
invariant mass, $W^2 > 4$ GeV$^2$, in the range $1<x<2$.
The notation for $\delta_{_T} f$ in Eq.\,(\ref{eqn:b1-parton})
indicates that $b_1$ probes very different spin structure in a hadron. 
Namely, $\delta_{_T} f$ is the {\it unpolarized} quark distribution
in a tensor-polarized hadron, whereas the polarized structure function
$g_1$ indicates the longitudinally polarized quark distribution
in a longitudinally polarized hadron.

A useful guideline for $b_1$ is expressed as the $b_1$ sum rule 
in the parton model \cite{b1-sum,tensor-pdfs}, and it is obtained
in the similar way to the Gottfried sum rule \cite{flavor3}:
\begin{align}
\! \! \!
\int dx \, b_1 (x) 
    & = - \lim_{t \to 0} \frac{5}{24} \, t \, F_Q (t) 
\nonumber \\
  & \ \ \ 
     + \frac{1}{9} \int dx
      \, \left [ \, 4 \, \delta_{_T} \bar u (x) +  \delta_{_T} \bar d (x) 
                     +   \delta_{_T} \bar s (x)  \, \right ] ,
\nonumber \\
 \int \frac{dx}{x}
 \, [F_2^p & (x) - F_2^n (x) ] 
   =  \frac{1}{3} 
   +\frac{2}{3} \int dx \, [ \bar u(x) - \bar d(x) ] .
\label{eqn:b1-sum-gottfried}
\end{align}
Here, the function $F_Q(t)$ is the electric quadrupole form factor
for the spin-1 hadron, so that the first term vanishes:
$\lim_{t \to 0} \frac{5}{24} t F_Q (t)=0$, whereas
the first term of the Gottfried sum is finite (1/3).
These sums originate from the fact that the valence-quark numbers depend
on the quark flavor but not on the hadron spin. 
As the Gottfried-sum-rule violation indicated the flavor asymmetric
distribution $\bar u (x)-\bar d(x)$, the $b_1$ sum-rule violation
could initiate the studies of finite tensor-polarized antiquark 
distributions. In fact, the HERMES data in the range $Q^2 > 1$ GeV$^2$
indicated that it is violated, 
$\int_{0.02}^{0.85} dx b_1(x) 
  = [0.35 \pm (\text{stat}) \pm 0.18 (\text{syst})] \times 10^{-2}$.
This suggestion of finite tensor-polarized antiquark distributions
should be tested by polarized proton-deuteron Drell-Yan process
at Fermilab or other hadron facilities. Recently, the tensor-polarization
asymmetry is theoretically estimated for the Fermilab Drell-Yan 
experiment \cite{Fermilab-dy}, and it is considered within 
the Fermilab E1039 experiment.

\noindent
{\bf Other definition of \boldmath{$b_{1-4}$}}:
There is another definition of the tensor-polarized structure functions 
by Edelmann, Piller, and Weise (EPW) \cite{epw-1997}
and it should not be confused with the one in Eq.\,(\ref{eqn:w-1}),
which is so far used in this article and for showing numerical results,
by Hoodbhoy, Jaffe, and Manohar (HJM) \cite{hjm89}.
They introduced structure functions which we denote
as $b_{1-3}^{\text{EPW}}$ and $\Delta^{\text{EPW}}$, which are
not equal to $b_{1-4}$.
Comparing the two hadronic tensors and structure functions functions,
we can relate the two sets to each other.  
For example, the HJM structure function $b_1$ so far used in our article 
is related to the EPW functions by
\begin{equation}
\left ( 1+\frac{Q^2}{\nu^2}\right ) b_1 
   =b_1^{\text{EPW}} -\frac{\Delta^{\text{EPW}}}{2}. 
\end{equation}
The structure functions $b_{1,2}$ in the two sets become equal 
in the scaling limit, and the higher-twist ones are considered 
equal to zero in the same limit.
The function $b_1^{\text{EPW}}$ is defined such that
the ratio of the following transverse structure functions,
$F_{UT_{LL},T}$ and $F_{UU,T}$,
defined in Eq.\,(\ref{eq:cross})
are given by the ratio of $b_1^{\text{EPW}}$ and $F_1$
as shown later in Eq.\,(\ref{eq:b1-epw-ratio}).

\section{Theoretical formalisms for $\bm{b}_{\bm{1}}$}
\label{formalism}

\subsection{Theory 1: Basic convolution description}
\label{convolution}

\begin{figure}[ht]
\begin{center}
   \includegraphics[width=6.5cm]{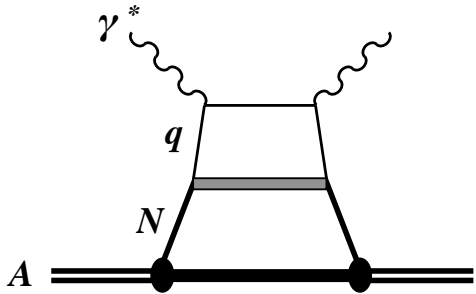}
\end{center}
\vspace{-0.5cm}
\caption{Convolution description for a structure function
of the deuteron. The notations $\gamma^*$, $q$, $N$, and $A$
indicate the virtual photon, quark, nucleon, and nucleus,
respectively.}
\label{fig:convolution-d}
\end{figure}

The most standard way of calculating nuclear structure functions
at medium- and large- $x$ regions ($x>0.2$)
is to use a convolution formalism. A nuclear hadron tensor 
$W_{\mu\nu}^A$ is given by the nucleonic one
$W_{\mu\nu}$ convoluted with a nucleon momentum distribution $S(p)$, 
which is called the spectral function, in a nucleus 
\cite{basic-conv,ek-2003,nuclear-sfs}:
\begin{equation}
W_{\mu\nu}^A (P_A, q) = \int d^4 p \, S(p) \, W_{\mu\nu} (p, q) ,
\label{eqn:w-convolution-w}
\end{equation}
where $p$ and $P_A$ are momenta for the nucleon and nucleus,
respectively. 
As illustrated in Fig.\,\ref{fig:convolution-d}, 
a nucleon is distributed in a nucleus by the spectral function
$S(p)$ and a quark is distributed in the nucleon by
the distribution function $q(x)$, and the overall
nuclear quark distribution is given by the convolution integral
of their functions. The spectral function is given by
\begin{equation}
S (p)= \frac{1}{A} \sum_i
      | \phi_i (\vec p \,) |^2 
     \delta \left ( p^0-M_A + \sqrt{M_{A-i}^{\ 2} 
                    + \vec p^{\ 2}} \, \right ) ,
\label{eqn:sp}
\end{equation}
in a simple shell model.
Here, $M_{A-i}$ is the mass of residual one-hole state by removing a nucleon,
and $\phi_i (\vec p \,)$ is the wave function of the nucleon.
The separation energy $\varepsilon_i$ is
the energy required to remove a nucleon from the state $i$,
and it is expressed by
the nuclear mass $M_A$ and the mass $M_{A-i}$ as
\begin{equation}
\varepsilon_i = (M_{A-i}+M_N) - M_A .
\end{equation}

In our actual calculation for the deuteron, a nonrelativistic 
relation is used for $\sqrt{M_{A-i}^{\ 2}+ \vec p^{\ 2}}$, so that 
the energy conservation by the $\delta$ function indicates
\begin{equation}
p^0= M_N - \varepsilon - \frac{\vec p^{\ 2}}{2 M_N},
\label{eqn:pn0}
\end{equation}
where $\varepsilon$ is the separation energy for the deuteron.
Since the large-momentum contribution decreases significantly
due to the the deuteron wave function $\phi (p)$, this nonrelativistic
approximation does not change the result to a significant amount.

As discussed in Ref.\,\cite{hjm89},
the hadron tensors could be expressed in terms of their helicity amplitudes
of the virtual photon as
\begin{align}
A_{hH,hH} (x,Q^2)  
   =  \varepsilon_h^{*\mu} \varepsilon_h^{\nu} \, 
       W_{\mu\nu}^D (p_{_D},q),
\label{eqn:A-helicity}
\end{align}
for the deuteron and the corresponding one $\hat A_{hs,hs} (x,Q^2)$ 
for the nucleon. The photon polarization vector 
$\varepsilon_{h}^{\,\mu}$ is
given by
\begin{align}
  \varepsilon_{h= \pm 1}^{\,\mu} 
     & = \frac{1}{\sqrt{2}} \, (0,\, \mp 1,\,  -i,\, 0), 
\nonumber \\
  \varepsilon_{h= 0}^{\,\mu} 
     & = \frac{1}{\sqrt{Q^2}} \, (|\vec q \,|,\, 0,\, 0,\, q^0).
\end{align}
Then, the helicity amplitudes are related to the structure function
$b_1$ of the deuteron and $F_1$ of the nucleon (denoted as $F_1^N$)
by the relations \cite{hjm89,kk08}
\begin{align}
b_1 & = A_{+0,+0} - \frac{A_{++,++} +A_{+-,+-}}{2} 
   \bigg |_{\text{LT}} ,
\nonumber \\
F_1^N & = \frac{ A_{+\uparrow,+\uparrow} +A_{+\downarrow,+\downarrow} }{2} .
\label{eqn:A-b1F1g1}
\end{align}
We note that the above relation for $b_1$ is the leading-twist (LT) 
expression which is strict only in the scaling limit.
We use these relations for estimating the structure function $b_1$.
At this stage, even leading theoretical calculations without 
higher-twist effects are valuable in comparison with the existing 
HERMES measurements.

Using these equation, we obtain the convolution expression 
for the structure function $b_1$ of the deuteron as
\begin{align}
b_1 (x,Q^2) & = \int \frac{dy}{y} 
\, \delta_T f(y) \, F_1^N (x/y,Q^2), 
\nonumber \\
\delta_T f(y) & \equiv f^0 (y)  - \frac{f^+ (y) + f^- (y)}{2} .
\label{eqn:b1-convolution}
\end{align}
Here, the structure function $b_1$ is defined by the one
per nucleon, the lightcone momentum distribution is expressed 
by the momentum-space wave function of the deuteron 
$\phi^H (\vec p \,)$ as
\begin{align}
f^H (y) = \int d^3 p \, y \, | \, \phi^H (\vec p \,) \, |^2
          \, \delta \left ( y - \frac{E-p_z}{M_N}   \right ) ,
\label{eqn:deuteron-momentum}
\end{align}
where the variable $y$ is the momentum fraction defined by
\vspace{-0.1cm}
\begin{equation}
y   =    \frac{M \, p \cdot q}{M_{N} \, P \cdot q} 
  \simeq \frac{2 \, p^-}{P^-} ,
\vspace{-0.1cm}
\end{equation}
where $p^-$ is a light cone momentum
[$\, p^- \equiv (p^0 -p^3)/\sqrt{2} \,$].
We consider a collinear frame with the photon three-momentum 
along the positive $z$-axis, and consequently the minus component 
($p^-$) of nucleon and deuteron momenta survives
as their light-front momentum component, whereas
it is $p^+$ if the $z$-axis is taken as the nucleon momentum
The function $F_1^N$ is defined for the nucleon 
by the average of the proton and neutron
structure functions: $F_1^N = (F_1^p + F_1^n)/2$.
Since $b_1$ is associated with unpolarized quark distributions,
the nucleon spin also does not appear in the convolution integral.
Namely, the momentum distribution is for the unpolarized nucleon:
$f^H (y) \equiv f_\uparrow^H (y) + f_\downarrow^H (y)$
if $f_s^H (y)$ indicates the distribution of the nucleon 
with the spin state $s$ and the deuteron spin $H$ 
along the $z$ axis.

The wave function of the deuteron is written as
\begin{align}
\phi^H  & (\vec p \,)  =  \phi_0 (p) \, Y_{00} (\hat p) \, \chi_{_H}
\nonumber \\
     &   \ \ \ \ 
     +  \sum_{m_L} \langle \, 2 m_L: 1 m_S \, | \, 1 H \, \rangle
            \, \phi_2 (p) \, Y_{2m_L} (\hat p) \, \chi_{m_S} ,
\label{eqn:p-wave-function}
\end{align}
where 
$\phi_0 (p)$ and $\phi_2 (p)$ are S- and D-state wave functions
with the D-state admixture probability $\int dp \, p^2 \, |\phi_2 (p)|^2$.
Here, $Y_{L m_L}$ is the spherical harmonic,
$\langle \, L m_L: S m_S \, | \, 1 H \, \rangle$
is the Clebsch-Gordan coefficient,
and $\chi_{m_S}$ is the spin wave function.
The coordinate-space wave function is written as
\begin{align}
\psi^H (\vec r \,) & =  [u_0 (r)/r ] \, Y_{00} (\hat r) \, \chi_{_H}
\nonumber \\
     &  \! \! \! \! \! \! \! \! \! \!
  +  \sum_{m_L} \langle \, 2 m_L: 1 m_S \, | \, 1 H \, \rangle
            \, [u_2 (r)/r] \, Y_{2m_L} (\hat r) \, \chi_{m_S} .
\label{eqn:r-wave-function}
\end{align}
Then, the momentum-space wave function is related to 
the coordinate-space one by
$\phi_L (p) = 4 \, \pi \, i^L \int dr \, r \, j_L (pr) \, u_L (r)$,
where $j_L (pr)$ is the spherical Bessel function. 
One may note that the D-state wave function has the negative sign
($\phi_2 (p)<0$) due to the $i^L$ factor although a different 
convention ($\phi_2 (p) \to - \phi_2 (p)$, namely without the $i^2$ factor)
is sometimes used for the D-state wave function.
Using the wave function in Eq.\,(\ref{eqn:p-wave-function})
for calculating the momentum distribution of 
Eq.\,(\ref{eqn:deuteron-momentum}), we obtain the tensor distribution
of Eq.\,(\ref{eqn:b1-convolution}) as
\begin{align}
\delta_T f(y)  = \int d^3 p \, y &
 \left [ - \frac{3}{4 \sqrt{2} \pi} \phi_0 (p) \phi_2 (p) 
  + \frac{3}{16\pi} |\phi_2 (p)|^2 \right ]
 \nonumber \\
 & \times
 (3 \cos^2 \theta -1) \, \delta \left ( y - \frac{p\cdot q}{M_N \nu}  \right ) .
\label{eqn:delta-t-f}
\end{align}
For normalizing the momentum distribution, we use the condition of
the baryon-number conservation,
$\int dy \, f^H (y) = \int d^3 p \, y \, |\phi^H (\vec p \,)|^2 =1$
\cite{basic-conv,ek-2003,nuclear-sfs}, which is slightly different 
from the nonrelativistic wave-function normalization without the $y$ factor.
This issue was discussed in the convolution description 
for nuclear structure functions,
so that the interested reader may look at the articles 
in Refs.\,\cite{basic-conv,ek-2003,nuclear-sfs}.
The expression of Eq.\,(\ref{eqn:delta-t-f}) is similar to
the one given in Ref.\,\cite{kh91}, which is the updated version of
the original convolution formalism in Ref.\,\cite{hjm89}.

The purposes of our work are to study the convolution expression 
independent from Ref.\,\cite{kh91} and to check their numerical result.
On the second point, we find very large differences from their $b_1$
as discussed in Sec.\,\ref{results}.
On the formalism, there are some differences.
First, our wave function is normalized by the baryon-number conservation
including the factor $y$ as it is usually used in the convolution 
formalism for the nuclear structure functions 
\cite{basic-conv,ek-2003,nuclear-sfs}. 
In Refs.\cite{hjm89,kh91}, the wave function is normalized by 
$\int dy f(y)=1$ as written below Eq.\,(22) of the Hoodbhoy-Jaffe-Manohar 
paper, in a similar way to satisfy the the baryon-number conservation
in this work, however, by using a relativistic correction factor
$1+\alpha_3$ with the Dirac spinor.
Second, $p^0$ is defined in the spectral function
with the energy-conserving $\delta$ function of Eq.\,(\ref{eqn:sp}),
which leads to the relation Eq.\,(\ref{eqn:pn0}) by the nonrelativistic
approximation. However, it is simply assumed as 
$p^0= M - \varepsilon + \vec p^{\ 2}/(2 M_N)$,
where the last kinetic term has the opposite sign,
in Refs.\cite{hjm89,kh91}.

For the $F_1^N$ structure function, we use the leading-order (LO) expression
with the longitudinal-transverse ratio 
$R=[ (1+Q^2/\nu^2) F_2^N -2x F_1^N]/(2xF_1^N)$ as
\begin{align}
F_1^N (x,Q^2) & = 
\frac{1+4 \, M_N^2 \, x^2/Q^2}{2 \, x \, [1+R(x,Q^2)]}
     \, F_2^N (x,Q^2) ,
\nonumber \\
F_2^N (x,Q^2)_{\text{LO}} & = x \sum_i e_i^2 
     \left [ q_i (x,Q^2) + \bar q_i (x,Q^2) \right ]_{\text{LO}} .
\label{eqn:f1-lo}
\end{align}
There exists a parton-model expression for $F_1^N$ by the Callan-Gross relation
to $F_2^N$ ($F_1^N=F_2^N /(2x)$) in the Bjorken-scaling limit. 
However, Eq.\,(\ref{eqn:f1-lo}) is practically used for calculating $F_1^N$ 
by taking into account the finite longitudinal-transverse ratio.
The structure function $F_1^N$ is for the nucleon within the deuteron.
We calculated it by neglecting nuclear corrections by the following reasons.
First, nuclear modifications are typically within a few percent 
in $F_2$ for the deuteron \cite{hkn07}. Furthermore, there is no experimental
signature on nuclear modifications of $R$ \cite{hermes-r},
although such effects could exist theoretically, for example,
by the Fermi motion of nucleons \cite{ek-2003}. 
In any case, nuclear medium effects are considered to be small
in the deuteron, so that they are neglected in our numerical estimates.

Using Eqs. (\ref{eqn:b1-convolution}), (\ref{eqn:delta-t-f}), and
(\ref{eqn:f1-lo}), we obtain numerical results for this theoretical model
as discussed in Sec.\,\ref{results}.
We should note that the leading-twist relation of 
Eq.\,(\ref{eqn:A-b1F1g1}) is used for obtaining the convolution 
equation (\ref{eqn:b1-convolution}), so that its numerical results 
are not precise at small $Q^2$. As for the structure function $F_1^N$ 
in the convolution integral, we used a realistic one, 
which corresponds most closely to the one obtained from experiments,
in Eq.\,(\ref{eqn:f1-lo}). This choice is also
intended for comparison with theory-2 results as mentioned in the end of
Sec.\,\ref{vna-model}. Therefore, higher-twist effects are contained
in the nucleonic structure-function level, whereas they are neglected
in the convolution expression in the theory-1 description.
We need to be aware it in looking at numerical results in 
Sec.\,\ref{results}.

\subsection{Theory 2: Virtual nucleon approximation}
\label{vna-model}

Next, we explain another convolution formalism by using the virtual nucleon
approximation. Before stepping into the model, we introduce a general formalism
for polarization factors. The density matrix for a spin-1 hadron is written 
by the spin-polarization vector $\vec {\mathcal P}$
and rank-2 spin tensor $T_{ij}$ as
\cite{Leader:2001gr}
\begin{align}
\rho = \frac{1}{3} \left [  \,
        1 + \frac{3}{2} \vec {\mathcal P} \cdot \vec S 
        + \sqrt{\frac{3}{2}} \, T_{ij} \,
        (S_i S_j + S_j S_i) \, \right ],
\label{eq:density-matrix}
\end{align}
where $\vec S$ is the $3 \times 3$ matrix representing
the spin operator $\hat {\vec S}$ for the spin-1 hadron.
The polarization vector $\vec {\mathcal P}$ and the rank-2 spin
tensor $T_{ij}$ are defined by
\begin{align}
\! \! \!
\vec {\mathcal P} = \langle \, \hat {\vec S} \, \rangle , \ \ \ 
T_{ij} = \frac{1}{2} \sqrt{\frac{3}{2}}
   \left (  \langle \, \hat S_i \hat S_j +  \hat S_j \hat S_i \, \rangle 
       - \frac{4}{3} \, \delta_{ij} \right ) .
\label{eq:p-tij}
\end{align}
The degrees of vector and tensor polarizations are given by
$\mathcal{P}=\sqrt{\vec{\mathcal P}^2}$ and
$T=\sqrt{\sum_{i,j} (T_{i,j})^2}$.
If the probabilities of spin states $+1$, $0$, and $-1$ are denoted
as $p^{+1}$, $p^0$, and $p^{-1}$, respectively, 
by taking the $z'$-axis as the quantization axis, 
the vector and tensor polarizations are 
\begin{align}
&{\mathcal P}_{z'} = p^+ - p^-,
&T_{z'z'} =  \frac{1}{\sqrt{6}} ( 1- 3 \, p^0). 
\end{align}
respectively.
We denote this tensor polarization also as
\begin{equation} \label{eq:Tzz}
\widetilde{T}_{\parallel\parallel}= \frac{1}{\sqrt{6}} (1-3p^0).
\end{equation}

The inclusive cross section of a charged-lepton deep inelastic 
scattering from a spin-1 target is generally expressed as
\begin{align}
\frac{d\sigma}{dx \, dQ^2} 
= \frac{\pi y^2\alpha^2}{Q^4(1-\epsilon)} &
\bigg[ F_{UU,T}+\epsilon F_{UU,L}
\nonumber \\
& \! \! \! \! \! \!  
+T_{\parallel\parallel} 
      \left( F_{UT_{LL},T}+\epsilon F_{UT_{LL},L} \right) 
\nonumber  \\ 
& \! \! \! \! \! \!  
+ T_{\parallel\perp} \cos\phi_{T_\parallel} \sqrt{2\epsilon(1+\epsilon)} \,
   F_{UT_{LT}}^{\cos\phi_{T_\parallel}} 
\nonumber  \\ 
& \! \! \! \! \! \!  
+ T_{\perp\perp} \cos(2\phi_{T_\perp}) \, \epsilon \,
   F_{UT_{TT}}^{\cos(2\phi_{T_\perp})}\bigg ]\,,
\label{eq:cross}
\end{align}
in terms of the spin-dependent factors and structure functions \cite{tagged-spin-1}.
Here, the $z$-axis is taken along the virtual-photon momentum 
direction ($\vec q \, /|\vec q \, |$). Then, the polarization factors 
$T_{\parallel\parallel}$, $T_{\parallel\perp}$, and $T_{\perp\perp}$
are related to $T_{ij}$ by the relations
$T_{\parallel\parallel}=T_{zz}$,
$T_{\parallel\perp} \cos\phi_{T_\parallel}=T_{xz}$, and 
$T_{\perp\perp} \cos(2\phi_{T_\perp})=T_{xx}-T_{yy}$
by assigning the angles $\phi_{T_\parallel}$
and $\phi_{T_\perp}$.
Namely, the tensor $T$ is decomposed in three parts: 
a projection on the longitudinal direction ($T_{\parallel\parallel}$), a 
projection on the transverse space ($T_{\perp\perp}$) and a mixed projection 
($T_{\parallel\perp}$), where longitudinal and transverse are relative to 
$\vec{q}$.
The angle $\phi_{T_\parallel}$ is the 
azimuthal angle of the transverse part of the mixed projection, and
the angle $\phi_{T_\perp}$ is the 
azimuthal angle in the transverse space of the projection.
If the deuteron is polarized along the virtual photon direction, only 
$T_{\parallel\parallel}$ is nonzero and given by Eq.~(\ref{eq:Tzz}).
If the deuteron is polarized along the lepton-beam axis, we have
$\phi_{T_\parallel}=\phi_{T_\perp}=0$, and
the remaining polarization factors in Eq.~(\ref{eq:cross}) can be related to 
$\widetilde T_{\parallel\parallel}$ of Eq.~(\ref{eq:Tzz}) through the 
transformation properties 
of the density matrix under rotations as follows:
\begin{align}
T_{\parallel\parallel} & =\frac{1}{4} \, [ 1+3\cos ( 2\theta_q ) ] \,
        \widetilde{T}_{\parallel\parallel} , \ \ \ 
T_{\parallel\perp} =\frac{3}{4} \sin (2\theta_q ) \, 
        \widetilde{T}_{\parallel\parallel} ,
\nonumber \\        
T_{\perp\perp} & =\frac{3}{4} \,
         [ 1-\cos (2\theta_q ) ] \, \widetilde{T}_{\parallel\parallel} ,
\end{align}
where $\theta_q$ is the angle between the lepton-beam ($z'$) 
and virtual-photon ($z$) direction. 
The variables $y$ and $\gamma$ are defined 
by the spin-1 hadron momentum $P$, its mass $M$, the initial
lepton momentum $\ell$, the momentum transfer $q$, and $Q^2$ as
\begin{align}
y=\frac{P\cdot q}{P \cdot \ell} , \ \ \ 
\gamma = \frac{\sqrt{Q^2}}{\nu}=\sqrt{1-\kappa}
\label{eq:xygamma}
\end{align}
The factor $\epsilon$ indicates the degree of
the longitudinal polarization of the virtual photon
as it appears in front of the longitudinal structure function $F_{UU,L}$,
and it is given by
\begin{align}
\epsilon 
= \frac{1}{ 1 + (1+\nu^2/Q^2) \tan^2 (\theta/2) } ,
\end{align}
where $\theta$ is the scattering angle of the charged lepton.
The six structure functions in Eq.~(\ref{eq:cross}) can be written  
by the virtual photon helicity amplitudes of the hadronic tensor 
in Eq.\,(\ref{eqn:A-helicity}). 
Then, the tensor polarized structure functions, 
which are used to calculate $b_1$ below, 
are expressed by the photon helicity amplitudes as
\begin{align}
& F_{UT_{LL},L}=\frac{4}{\sqrt{6}}
\left(A_{+0,+0}-A_{00,00}\right) ,
\nonumber\\
&F_{UT_{LL},T}=\frac{2}{\sqrt{6}}
\left(A_{++,++}-2A_{+0,+0}+A_{+-,+-}\right) ,
\nonumber\\
&F_{UT_{LT}}^{\cos\phi_{T_\parallel}}=-\frac{4}{\sqrt{6}}
\, \Re e \left( A_{+0,0+} -A_{+-,00} \right) ,
\nonumber\\
&F_{UT_{TT}}^{\cos(2\phi_{T_\perp})}=-\sqrt{\frac{2}{3}}
\, \Re e A_{+-,-+} \ .
\label{eqn:futll-lt}
\end{align}

Using the expression of Eq.\,(\ref{eqn:w-1}) for the hadron tensor 
in terms of the polarized structure functions $b_{1-4}$
and the helicity amplitude definition of Eq.\,(\ref{eqn:A-helicity}), 
we obtain \cite{tagged-spin-1}
\begin{align}
& \! \! \!
F_{UT_{LL},L}   = 
\frac{1}{x_D}\sqrt{\frac{2}{3}} \bigg[ 2(1+\gamma^2)x_D b_1
-(1+\gamma^2)^2 \left(\frac { 1 } { 3}b_2+b_3+b_4\right) 
\nonumber\\ 
& \ \ \ \ \ \ \ \ \ \ \ \ \ \ \ \ \ \ \ \ \ \ \,
\left.-(1+\gamma^2)\left(\frac{1}{3}b_2-b_4\right)
-\left(\frac{1}{3}b_2-b_3 \right)\right ] ,
\nonumber\\
& \! \! \!
 F_{UT_{LL},T}   = 
-\frac{1}{x_D}\sqrt{\frac{2}{3}}\left[2(1+\gamma^2)xb
_1-\gamma^2\left(\frac{1}{6}
b_2-\frac{1}{2}b_3\right)\right ] ,
\nonumber\\
& \! \! \! 
 F_{UT_{LT}}^{\cos\phi_{T_\parallel}} 
 = -\sqrt{\frac{2}{3}}\frac{\gamma}{2x_D}\left[(1+\gamma^2)
    \left(\frac{1}{3} b_2-b_4\right)+\left(\frac{2}{3}b_2-2b_3\right)\right] ,
\nonumber\\
& \! \! \! 
 F_{UT_{TT}}^{\cos (2\phi_{T_\perp})} 
  = -\sqrt{\frac{2}{3}}\frac{\gamma^2}{x_D} 
\left(\frac{1}{6}b_2-\frac{1}{2}b_3 \right) .
\label{eq:inclfinal}
\end{align}
Therefore, the $b_1$ is written through the structure functions 
$F_{UT_{LL},T}$ and $F_{UT_{TT}}$ as
\begin{equation} 
b_1=- \frac{1}{1+\gamma^2} \sqrt{\frac{3}{8}} \,
   \bigg [ F_{UT_{LL},T}
    +F_{UT_{TT}}^ { \cos(2\phi_ { T_\perp }) } \bigg ] .
\label{eq:b1_f}
\end{equation}

We also show the relation between these structure functions 
and the EPW function $b_1^{\text{EPW}}$, which is related
by the ratio of transverse structure functions as
\begin{align}
\sqrt{\frac{2}{3}}\frac{F_{UT_{LL},T} }{F_{UU,T} }
&=\frac{A_{++,++}-2A_{+0,+0}+A_{+-,+-}}{A_{++,++}+A_{+0,+0}+A_{+-,+-}}
\nonumber \\
&= 
-\frac{2}{3}\frac{b_1^{\text{EPW}}}{F_1}
\label{eq:b1-epw-ratio} .
\end{align}
This equality is not valid with the HJM $b_1$ because
the structure functions $b_2,b_3$
also contribute to $F_{UT_{LL},T}$
as shown in Eq.\,(\ref{eq:inclfinal}).

\begin{figure}[t]
 \begin{center}
  \includegraphics[width=0.25\textwidth]{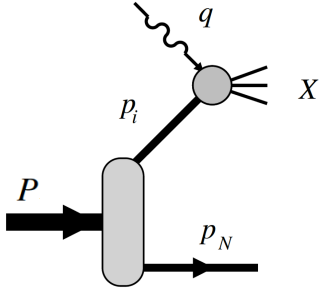}
 \end{center}
\vspace{-0.4cm}
\caption{
\label{fig:VNA_IA} 
Impulse approximation diagram in the VNA. 
For the inclusive reaction, we integrate over the phase space 
of the spectator nucleon.}
\end{figure}

Next, we explain how to calculate the structure functions 
$F_{UT_{LL},T}$ and $F_{UT_{TT}}$ 
for the deuteron by the virtual nucleon approximation (VNA),
which considers the $np$ component of the light-front deuteron wave function.
As shown in Fig.~\ref{fig:VNA_IA}, the virtual photon interacts 
with one nucleon which is off the mass shell in the DIS reaction,
while the second noninteracting ``spectator'' is assumed to be
on its mass shell. Then, the inclusive structure functions 
in the impulse approximation are obtained by integrating 
over all possible spectator momenta $\vec p_N$. 

In the following, we explain the outline for deriving 
the tensor polarized structure functions in the light-front 
formulation of the VNA. In Fig.~\ref{fig:VNA_IA},
$P$, $p_i$, and $p_N$ are momenta for the deuteron ($P=p_i + p_N$), 
the struck nucleon, and the on shell spectator, respectively.
The convolution approach for the symmetric part of 
the hadron tensor is given in the VNA model for the deuteron as 
\cite{tagged-spin-1}
\begin{equation}
\! 
W_{\mu\nu}^{\lambda' \lambda} (P,q)
 \! = \! 4(2\pi)^3 \! \! \int \! d\Gamma_N
\frac{\alpha_{_N}}{\alpha_i} 
W^N_{\mu\nu} (p_{i},q)
\rho_D(\lambda',\lambda) ,
\label{eq:htensorfinal}
\end{equation}
where $W^N_{\mu\nu}$ is the hadron tensor for the nucleon
and $d\Gamma_N$ is the Lorentz invariant phase space for the spectator nucleon.
We note that only the symmetric term of $W_{\mu\nu}^{\lambda' \lambda}$
under the exchange $\mu \leftrightarrow \nu$ is relevant for the 
tensor structure functions $b_{1-4}$.
The factor $4 (2\pi)^3$ arises in defining the deuteron light cone 
wave function, which is shown later in Eqs.\,(\ref{eqn:rho-d}) 
and (\ref{eq:D_lf_wf}), and the factor $\alpha_{_N}/\alpha_i$ appears because 
the hadron tensor $W_{\mu\nu}$ is for the nucleon with momentum $p_i$
instead of the nucleon at rest \cite{tagged-spin-1}.
Here,the light cone momentum fractions are defined for 
the interacting ($i$) and spectator ($N$) nucleons as
\begin{align}
\alpha_i=\frac{2 \, p_i^-}{P^-}\,, \ \ \ 
\alpha_{_N}=\frac{2 \, p_N^-}{P^-}=2-\alpha_i\, .
 \label{eq:lcfraction_Def}
\end{align}
Next, we define the relative momentum $\vec k$ of 
two nucleons by
\cite{relative-momentum}
\begin{align}
& k = \sqrt{E_k^2 -M_N^2}, \ \ \ 
E_k^2  = \frac{m_N^2+\vec k^{\perp 2}}{\alpha_i(2-\alpha_i)}, 
\nonumber \\
& k^3 = (1- \alpha_i )E_k , \ \ \ \,
\vec k^\perp = \vec p_i^\perp + \frac{\alpha_i}{2} \vec P^\perp .
\label{eq:lf_k_mom}
\end{align}
The momentum $k$ corresponds with the relative momentum 
of the free two nucleon state with identical light-front momentum 
components ($P^-$, $P^\perp$) as the deuteron,
and the overlap of this state with the deuteron 
defines the deuteron light-front wave function.
The spectator nucleon phase-space element can be written with the variables of 
Eqs.~(\ref{eq:lcfraction_Def}) and (\ref{eq:lf_k_mom}) as
\begin{align}
 d\Gamma_N = \frac{d^3 p_{_N}}{2E_{p_{_N}}(2\pi)^3} = \frac{d\alpha_id\vec 
p_i^\perp}{2\alpha_i(2\pi)^3}=\frac{\alpha_i d^3\vec k}{2E_k(2\pi)^3}.
\end{align}

In Eq.\,(\ref{eq:htensorfinal}), the deuteron density 
$\rho_D(\lambda',\lambda)$ is defined by
the light-front deuteron wave function 
$\Psi^{D}_\lambda(\vec{k},\lambda'_N,\lambda_N)$ 
\begin{equation}
\! 
\rho_D(\lambda',\lambda)
= \! \! \! \sum_{\lambda_{N},\,\lambda_{N}'} \! \! \!
\frac{[\Psi^{D}_{\lambda'}(\vec k,\lambda_{N}',\,\lambda_{N}) ]^\dagger
\Psi^{D}_\lambda(\vec k,\lambda_{N}',\,\lambda_{N})}{\alpha_{_N} \, \alpha_i} ,
\label{eqn:rho-d}
\end{equation}
and it is expressed as
\cite{tagged-spin-1,Frankfurt:1981mk,Keister:1991sb,cms-vna}
\begin{align}
\! \! \!
\Psi^D_\lambda  &  (\vec k,\lambda_1,\lambda_2) 
=\sqrt{E_{k}}\sum_{\lambda'_1,\lambda'_2}\mathcal{D}^{
\frac{1}{2}}_{\lambda_1\lambda'_1}[R \, (k_{1}/m_N)]
\nonumber \\ 
& 
\times
\mathcal{D}^{\frac{1}{2}}_{\lambda_2\lambda'_2}[R \, (k_{2}/m_N)]
\sum_{\substack{l=0,2\\\lambda_l\lambda_S}}
\langle l\lambda_l : S\lambda_S|j \lambda\rangle 
\nonumber \\ 
& \times
\langle s_1\lambda'_1 : s_2\lambda'_2|1\lambda_S\rangle 
Y_{l \lambda_l}(\Omega_{\bm k}) (-i)^l \phi_{l}(k) ,
\label{eq:D_lf_wf}
\end{align}
where $\mathcal{D}^{\frac{1}{2}}_{\lambda \lambda'}$ is 
the rotation matrix
and $\phi_i (k)$ is the deuteron wave function
with the orbital angular momentum $l$.
The notation $R$ indicates the Melosh rotation,
which relates canonical and light-front quantized spinors, 
the nucleon momenta [$k_1 =(E_k,\vec k)$, $k_2 =(E_k,-\vec k)$].
The wave functions are approximated by the nonrelativistic ones
$\phi_0 (k) = U(k)$ and $(-i)^2 \phi_2 (k) = W(k)$.
It is important to note, as discussed below Eq.\,(\ref{eqn:delta-t-f}),
that the deuteron light-front wave function of Eq.~(\ref{eq:D_lf_wf}) 
satisfies the baryon and momentum sum rules
\cite{tagged-spin-1,Frankfurt:1981mk}
\begin{align}
&\sum_{\lambda_1 , \, \lambda_2}\int \frac{d \alpha_i \, d \vec k^\perp}
{\alpha_i(2-\alpha_i)}\left|\Psi^{D}_\lambda(\alpha_i,
\vec k^\perp,\lambda_1,\lambda_2)\right|^2=1 ,
\nonumber \\ 
&\sum_{\lambda_1 , \, \lambda_2}\int \frac{d \alpha_i \, d \vec k^\perp}
{\alpha_i(2-\alpha_i)} \alpha_i\left|\Psi^{D}_\lambda
(\alpha_i,\vec k^\perp,\lambda_1,\lambda_2)\right|^2=1\,.
\end{align}

Using the convolution equation (\ref{eq:htensorfinal}) 
by the VNA model with the deuteron wave function (\ref{eq:D_lf_wf}),
the helicity amplitudes (\ref{eqn:A-helicity}),
and their relations to $F_{UT_{LL},L}$ and $F_{UT_{LL},T}$ 
in Eq.\,(\ref{eqn:futll-lt}),
we obtain the structure functions in the VNA convolution model as
\begin{widetext}
\begin{align}
F_{UT_{LL},T} &= - \int \frac{k^2}{\alpha_i} dk \, d(\cos\theta_k) 
\left[ F_{1}^N(x_i,Q^2) -\frac{T^2}{2 \, p_i \cdot q}F_{2}^N(x_i, Q^2) \right]
\nonumber \\
& \ \ \ \ \ \ \ \ \ \ \ \ \ \ \ \ \ \ \ \ \ \ \ \ \ \ \ \ \ \ 
\times
\sqrt{\frac{3}{2}}\left[\frac{U(k)W(k)}{\sqrt{2}}+\frac{W(k)^2}{4}
\right][3\cos (2\theta_k) +1] ,
\nonumber\\
F_{UT_{TT}}^{\cos(2\phi_{T_\perp})} &= - \int \frac{k^2}{\alpha_i} 
dk \, d(\cos\theta_k) \frac{-T^2}{2 \, p_i \cdot q}F_{2}^N(x_i,Q^2) 
\sqrt{\frac{3}{2}}\left[\frac{U(k)W(k)}{\sqrt{2}}
   +\frac{W(k)^2}{4}\right]\sin^2\theta_k ,
\label{eq:vna}
\end{align}
where the structure functions $F_1^N$ and $F_2^N$ are defined
by the averages of the proton and neutron functions as defined 
in Sec.\,\ref{convolution} and $\theta_k$ is the angle between $\vec k$ and 
$\vec q$.
Here, $T^\mu$ and $L^\mu$ are defined by
\begin{align}
T^\mu  = p_N^\mu+\frac{p_N\cdot q}{Q^2} q^\mu 
         - \frac{p_N\cdot L}{L^2} L^\mu , \ \ \ 
L^\mu=P^\mu+\frac{P\cdot q}{Q^2} q^\mu .
\end{align}
The nucleon structure functions $F_{1}^N$ and $F_{2}^N$ are evaluated at 
$ x_i={Q^2}/{(2 p_i \cdot q)}\simeq x / \alpha_i $. 
Substituting the structure functions of Eq.\,(\ref{eq:vna}) 
into Eq.\,(\ref{eq:b1_f}), we finally obtain the expression 
for $b_1$ in the VNA model,
\begin{align}
b_1(x,Q^2)=\frac{3}{4(1+\gamma^2)} \int \frac{k^2}{\alpha_i} 
dk \, d(\cos\theta_k) &
\left[ F_{1}^N(x_i,Q^2) \left(6\cos^2\theta_k-2\right)
-\frac{T^2 } {2 \, p_i \cdot q } F_ { 2 }^N (x_i ,Q^2)
\left(5\cos^2\theta_k-1\right) \right]
\nonumber \\
& \times
\left[ \frac{U(k)W(k)}{\sqrt{2}}+\frac{W( k)^2}{4}\right] .
\label{eq:b1_vna}
\end{align}
\end{widetext}
In deriving this expression of the theory 2, the Bjorken scaling limit is not
taken and the higher-twist effects are contained as it is clear by the additional 
term of $F_2^N$ in comparison with Eq.\,(\ref{eqn:b1-convolution}) 
of the theory 1. 
Because of the higher-twist effects included, for self-consistency, the theory 2 
should include nucleon structure functions that also contain higher-twist effects. 
For the purpose of the comparison with the theory 1 in evaluating only 
higher-twist effects originating from the nuclear part, we use the same 
nucleon structure function used in theory 1 [Eq.~(\ref{eqn:f1-lo})].

\subsection{Tensor-polarization asymmetry $A_{zz}$ and structure function $b_1$}
\label{Azz-b1}

In the unpolarized charged-lepton DIS from the polarized deuteron like 
the HERMES experiment \cite{hermes05}, the cross section 
with target polarization along the beam direction 
is written as 
\begin{equation}
 \frac{d\sigma}{dxdQ^2}=\frac{d\sigma^U}{dxdQ^2}\left(
1+\frac{1}{2}P_{zz}A_{zz}\right) ,
\label{eq:exp_cross}
\end{equation}
where $d\sigma^U / dx dQ^2$ is the unpolarized cross section, and 
$P_{zz}$ is related to the density matrix variables 
defined in the beginning of Sec.\,\ref{vna-model} as
\begin{equation}
 P_{zz}=\sqrt{6}\, \widetilde{T}_{zz}=p^++p^--2p^0\,.
\end{equation}
Comparing Eq.~(\ref{eq:exp_cross}) with Eq.~(\ref{eq:cross}), we can write the 
tensor asymmetry $A_{zz}$ as
\begin{align}
A_{zz} & =\frac{2\sigma^+-2\sigma^0}{2\sigma^++\sigma^0}
=
\frac{\sqrt{2}}{4\sqrt{3} \left ( F_{UU,T}+\epsilon F_{UU,L} \right )}
\nonumber \\
& \times
\bigg\{
[1+3\cos(2\theta_q)] \left( 
F_{UT_{LL},T}+\epsilon 
F_{UT_{LL},L} \right)
\nonumber \\
& \ \ \ \ \ 
+3\sin(2\theta_q)
\sqrt{2\epsilon(1+\epsilon)}
F_{UT_{LT}}^{\cos\phi_{T_\parallel}}
\nonumber \\
& \ \ \ \ \ 
+3[1-\cos(2\theta_q)]\epsilon 
F_{UT_{TT}}^{\cos2\phi_{T_\perp}} \bigg\} ,
\label{eq:Azz_rot}
\end{align}
where $\sigma^i$ is the cross section with the target polarization $i$ 
along the beam ($z'$ axis) and we took $\sigma^+=\sigma^-$ because 
of the parity invariance.

In the HERMES analysis, $b_1$ was then extracted from $A_{zz}$ using
\begin{equation}
 A_{zz}=-\frac{2}{3}\frac{b_1}{F_1} .
 \label{eq:b1wrong}
\end{equation}
This equation is correct as an equality if the following 
two conditions are satisfied.
\begin{itemize}
 \item[(1)] The deuteron is polarized along the photon direction, namely $\theta_q=0$.
 \item[(2)] The Bjorken scaling limit ($Q^2 \rightarrow \infty, x\, \text{finite}, 
\gamma \rightarrow 0$) is taken.
It implies the Callan-Gross relations for the structure functions 
($2x_D F_1 =F_2$, $2x_D b_1 =b_2$) and neglect of the higher-twist structure functions
$b_{3,4}$. 
\end{itemize}
This can be seen by putting $\theta_q=0$ in Eq.~(\ref{eq:Azz_rot}). 
The surviving structure functions in the scaling limit after 
applying the Callan-Gross relations become
\begin{align}
  &F_{UT_{LL},T}=-2\sqrt{\frac{2}{3}}b_1, &F_{UT_{LL},L}=0 \,,
  \nonumber\\
  &F_{UU,T}=2F_1, &F_{UU,L}=0\,,
\end{align}
which leads to Eq.~(\ref{eq:b1wrong}).
The theory 2 includes higher-twist corrections and can test the above assumptions.
According to our estimate, there are significant higher-twist effects, so that
the Callan-Gross relations are not satisfied and the functions $b_{3,4}$ are not
very small in comparison with the leading ones $b_{1,2}$ 
as shown in Table.\,\ref{tab:bstruc-1234}.
These observations and the value of $\gamma$ for the HERMES kinematics indicate
that including higher-twist effects might be needed for an improved extraction 
of $b_1$. 

\begin{widetext}
\begin{center}
\begin{table}[h!]
\caption {Theory-2 calculations of the four tensor-polarized structure functions
for kinematics of the HERMES $b_1$ data \cite{hermes05}.}
  \centering
  \begin{tabular}{|c|c|c|c|c|c|c|c|}
    \hline 
    $x$ & $Q^2$ (GeV$^2$) & $b_1$($10^{-4}$) & $b_2$($10^{-5}$) & $b_3$($10^{-3}$) 
                          & $b_4$($10^{-3}$) & $b_2/(2x_D b_1)$ & $\gamma$\\
    \hline
    0.012 & 0.51 & \ 2.81  & \ \ 0.264      & -1.34 & \ 5.06 & 0.783   & 0.0315 \\
    0.032 & 1.06 & \ 6.92  & \ \ 1.97 \     & -1.87 & \ 7.51 & 0.890   & 0.0583 \\
    0.063 & 1.65 & \ 3.50  & \ \ 0.265      & -2.02 & \ 7.96 & 0.120   & 0.0920 \\
    0.128 & 2.33 &  -1.80  &   \ -7.38 \ \  & -2.13 & \ 7.49 & 3.20 \  & 0.157 \ \\
    0.248 & 3.11 &  -8.39  &    -28.1 \ \ \ & -2.09 & \ 4.58 & 1.35 \  & 0.264 \ \\
    0.452 & 4.69 &  -6.18  &    -21.7 \ \ \ & -1.11 &  -0.58 & 0.777   & 0.392 \ \\
\hline
  \end{tabular}
\label{tab:bstruc-1234}
\end{table}
\end{center}
\end{widetext}

\section{Results}
\label{results}

In showing our numerical results on $b_1$ in the convolution picture,
we need to choose (1) deuteron wave function, (2) parton distribution
functions (PDFs), and (3) longitudinal-transverse structure function ratio.
Here, the CD-Bonn wave function is used for the deuteron \cite{cd-bonn-2001},
the MSTW2008 (Martin-Stirling-Thorne-Watt, 2008) 
leading-order (LO) parametrization 
for the PDFs \cite{MSTW2008}, and the SLAC-R1998 parametrization 
for the longitudinal-transverse ratio $R$ \cite{r1998}. 
We also tested other wave functions and parametrizations, 
but numerical results do not change by a significant amount.
There is a source of the uncertainty due to our knowledge of the high-momentum 
part of the deuteron wave function, and it reveals itself at $x>0.8$ kinematics.
The experimental separation energy of the deuteron 2.22457 MeV
\cite{d-energy} is used in our numerical evaluation.

\begin{figure}[b!]
\begin{center}
   \includegraphics[width=6.5cm]{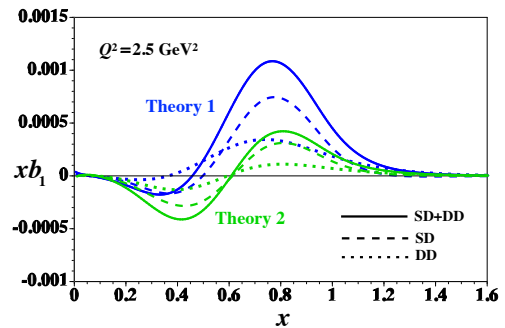}
\end{center}
\vspace{-0.5cm}
\caption{The structure function $b_1$ calculated 
by the two convolution descriptions of
Eqs.\,(\ref{eqn:b1-convolution}) and (\ref{eq:b1_vna})
at $Q^2$=2.5 GeV$^2$.
The dashed, dotted, and solid curves indicate contributions 
to $xb_1$ from the SD term, DD term, and their summation.
Two sets of theory curves are shown for the theory 1 and theory 2.}
\label{fig:xb1-sd-dd}
\end{figure}

In Fig.\,\ref{fig:xb1-sd-dd}, the calculated functions $x b_1$ are
shown for the SD interference term ($\propto\phi_0 \phi_2$), 
DD term ($\propto |\phi_2 |^2$), and their summation at $Q^2$=2.5 GeV$^2$ 
by using the two convolution descriptions in 
Eqs.\,(\ref{eqn:b1-convolution}) and (\ref{eq:b1_vna}).
This $Q^2$ scale is taken because of later comparison with the HERMES
data, where the $Q^2$ average is $Q^2$=2.5 GeV$^2$.
The SD contribution is larger than the DD one; however, the DD term is not
small as suggested by the magnitude of the D-state admixture of
4.85\% \cite{cd-bonn-2001}. It indicates that high-momentum components 
of the deuteron wave functions play an important role in the standard
convolution description for $b_1$.
Furthermore, the overall sign of the SD term is opposite to
the previous estimate in Ref.\,\cite{kh91}. 
Our convolution formalisms are similar to the one in Ref.\,\cite{kh91};
however, the SD term is completely different even in sign.
Since the SD contribution is the dominant term, this finding is 
important for future studies for an experimental comparison 
and in considering possible theoretical mechanisms of 
the tensor polarization in the parton level.

In addition, it is noteworthy to find the distribution at $x>1$,
whereas it vanishes according to the analysis of Ref.\,\cite{kh91}.
However, this region will be dominated by quasi-elastic scattering 
at moderate $Q^2$, and it will require the subtraction of this contribution
for the DIS analysis.
Since the Bjorken $x$ is defined by the same definition as
Eq.\,(\ref{eq:x}) in the convolution equations of Refs.\,\cite{hjm89,kh91},
although it seems to be defined by $x=Q^2/(2 M \nu)$
with the deuteron mass $M$ in the beginning of 
the Hoodbhoy-Jaffe-Manohar paper, the function $b_1$ should 
be finite even at $x>1$. In fact, the upper limit of their 
convolution integral is $y_{\text{max}}=2$. We could not figure out 
the reason why they do not have a $b_1$ distribution at $x>1$ 
in Ref.\,\cite{kh91}.

\begin{figure}[t!]
\begin{center}
   \includegraphics[width=6.5cm]{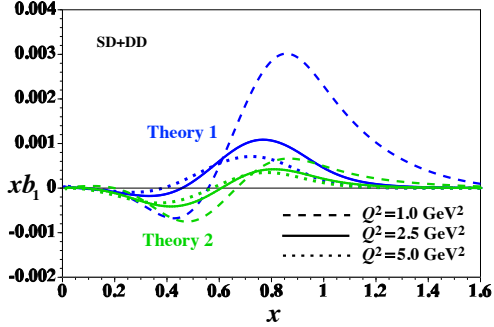}
\end{center}
\vspace{-0.5cm}
\caption{$Q^2$ dependence of the structure function $b_1$ 
by the two convolution descriptions of
Eqs.\,(\ref{eqn:b1-convolution}) and (\ref{eq:b1_vna})
at $Q^2$=1.0, 2.5, and 5.0 GeV$^2$.
The dashed, dotted, and solid curves indicate contributions 
to $xb_1$ from the SD term, DD term, and their summation.}
\label{fig:xb1-1-2_5-5}
\end{figure}

We used two theoretical models. They are similar but there are
some differences. First, the theory 2 in Eq.\,(\ref{eq:b1_vna})
includes other terms like $F_2$ in the convolution integral
as a higher-twist contribution. There are also differences
in kinematical treatments as shown in Eqs.\,(\ref{eqn:b1-convolution}) 
and (\ref{eq:b1_vna}). It suggests that $b_1$ is sensitive 
to dynamical details for describing the deuteron.
There exist significant differences between the two model predictions,
and their possible sources should be discussed.
First, the differences partly come from the higher-twist effects,
as it is clear from the large differences at $Q^2 =$1 GeV$^2$
and also from  Table.\,\ref{tab:bstruc-1234}.
To remove such effects, we took the scaling limit $\gamma \to 0$ 
in the theory 2 and both results become similar.
However, a complete agreement was not obtained even in this limit
and the remaining differences come from the slightly different 
normalizations and relativistic treatments 
for the deuteron wave functions.

\begin{figure}[b!]
\begin{center}
   \includegraphics[width=6.5cm]{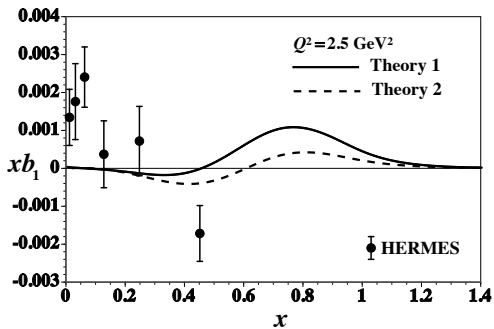}
\end{center}
\vspace{-0.5cm}
\caption{Calculated structure functions are compared with 
HERMES experimental data.
The solid and dashed curves indicate the functions 
$xb_1$ for theory 1 and theory 2, respectively, $Q^2=2.5$ GeV$^2$.
Here, the MSTW2008 PDFs are used as they are used
in Figs.\,\ref{fig:xb1-sd-dd} and \ref{fig:xb1-1-2_5-5}.}
\label{fig:xb1-hermes}
\end{figure}

The $Q^2$ dependence of $b_1$ is shown in Fig.\,\ref{fig:xb1-1-2_5-5}
by taking $Q^2$=1.0, 2.5, and 5.0 GeV$^2$ in the convolution models.
There are significant variations in $b_1$ in the region
1 GeV$^2 <Q^2<5$ GeV$^2$. This fact also indicates that 
$b_1$ is sensitive to dynamical aspects of hadron structure.
There are large differences between theory 1 and theory 2 
at $Q^2=1$ GeV$^2$, they are mainly due to the higher-twist effects 
which are significant at small $Q^2$.

\begin{figure}[b!]
\begin{center}
   \includegraphics[width=6.5cm]{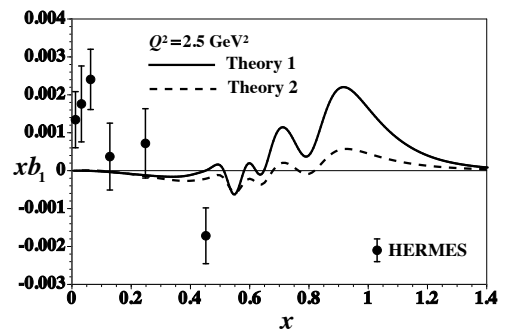}
\end{center}
\vspace{-0.5cm}
\caption{Calculated structure functions are compared with 
HERMES experimental data.
The solid and dashed curves indicate the functions 
$xb_1$ for theory 1 and theory 2, respectively, $Q^2=2.5$ GeV$^2$.
The Bodek {\it et al.} structure functions \cite{Bodek-etal-1979} 
are used instead of the MSTW2008 in Fig.\,\ref{fig:xb1-hermes}.
The other conditions are same as the ones in Fig.\,\ref{fig:xb1-hermes}.}
\label{fig:xb1-hermes-br}
\end{figure}

Next, we compare our total contribution from the SD and DD terms 
with the HERMES measurement on $b_1$. 
In Fig.\,\ref{fig:xb1-hermes}, our $x b_1$ curves are shown
at $Q^2$=2.5 GeV$^2$ for comparison with the HERMES data
because the average HERMES scale is $Q^2 =2.5$ GeV$^2$.
In general, the magnitude of $x b_1$ is much smaller than
the HERMES data at $x<0.5$, which means that 
the differences cannot be explained by the conventional deuteron
model and new hadron physics, at least beyond the current standard 
convolution description, is possibly needed for their interpretation.
Because the $b_1$ is sensitive to the D state and it is distributed
at relative large $x$, it is worthwhile to look at resonance effects
which could persist even in the deep inelastic region ($W^2 \ge 4$ GeV$^2$)
at $Q^2$ of a few GeV$^2$. In order to investigate such effects,
we take the structure-function parametrization of 
Bodek {\it et al.} in Ref.\,\cite{Bodek-etal-1979}
and show $b_1$ in Fig.\,\ref{fig:xb1-hermes-br}.
At $x>0.5$, an interesting bumpy resonance structure appears 
in $b_1$ according to the convolution picture.
As the quark-hadron duality indicates \cite{duality},
if the bumpy functions are averaged and they are approximated 
by smooth curves, they roughly agree with the $b_1$ functions
in Fig.\,\ref{fig:xb1-hermes}.

We reiterate the major points of our results.
\vspace{-0.20cm}
\begin{itemize}
 \item[(1)] Our convolution results for $b_1$ are numerically very different 
 from the ones in Ref.\,\cite{kh91}, especially in the SD contribution,
 although the theoretical formalisms are similar.
\vspace{-0.20cm}
 \item[(2)] There are finite distributions in $b_1$ even 
 at $x>1$, whereas there is no distribution in Ref.\,\cite{kh91}.
\vspace{-0.20cm}
 \item[(3)] Our convolution calculations for $b_1$ by the standard deuteron
 picture are very different from the HERMES measurement.
 It could suggest a new hadron-physics mechanism for interpreting
 the differences.
\end{itemize}
\vspace{-0.10cm}

Fortunately, a JLab experiment was approved for measuring $b_1$
accurately at medium $x$ \cite{Jlab-b1} and also
an experiment to measure the tensor-polarization asymmetry $A_{zz}$ 
at large $x$ is possible \cite{azz}, 
so that the situation should become much clearer in a few years.
There is also a possibility to measure the tensor-polarized
antiquark distributions in proton-deuteron Drell-Yan process
with the tensor-polarized deuteron target at Fermilab 
\cite{pd-drell-yan,ks-2016,Fermilab-dy}. Furthermore, it should be an interesting
topic to investigate $b_1$ at the future electron-ion-collider project
\cite{eic} and other hadron facilities 
such as Brookhaven National
Laboratory-RHIC, CERN-COMPASS, Japan Proton
Accelerator Research Complex \cite{j-parc}, Gesellschaft f\"ur
Schwerionenforschung-Facility for Antiproton and Ion
Research, and Institute for High Energy Physics in Russia.
The structure function $b_1$ could be also investigated at 
the International Linear Collider (ILC) if
a fixed-target experiment becomes possible in the similar way
with the TESLA-N project \cite{ilc}.

\section{Summary}\label{summary}

We calculated the tensor-polarized structure function $b_1$
for the spin-1 deuteron in the standard convolution description. 
The structure function $b_1$ is given by
the tensor-polarized light cone momentum distribution for the nucleon
convoluted with the unpolarized structure function of the nucleon. 
Two convolution models are used for evaluating $b_1$.
One is a basic convolution model and the other is the VNA model.
Our numerical results indicate that these standard theoretical
predications are much different from the HERMES measurements.
Furthermore, significantly large distributions are predicted,
at large $x$ ($x>0.8$) and even at extremely large $x$ ($x>1$). 
Since our results are very different from the HERMES measurement, 
new hadronic mechanisms could be needed for interpreting the data
although there is still some room to improve the differences
due to the higher-twist effects and the experimental extraction
of $b_1$ from $A_{zz}$.
The HERMES data have large uncertainties; however, 
upcoming JLab experimental measurements will improve 
on the size of the errors.
In addition, there are experimental possibilities at Fermilab,
EIC, and other facilities to investigate the tensor-polarized
structure functions. It is now a good opportunity to understand
the tensor structure in terms of quark and gluon degrees of 
freedom. Obviously, we need further theoretical studies 
on $b_1$ and other spin-1 structure functions possibly by
including exotic mechanisms.

\begin{acknowledgments}

This work was supported by Japan Society for the Promotion of Science (JSPS)
Grants-in-Aid for Scientific Research (KAKENHI) Grant No. JP25105010.
It is also partly supported by the National Natural Science Foundation 
of China under Grant No.~11475192 and the fund by the Sino-German 
CRC 110 ``Symmetries and the Emergence of Structure in QCD" project 
(NSFC Grant No. 11621131001). YBD (SK) thanks warm hospitality of 
the KEK theory center (IHEP, FIU) during his visit.
\end{acknowledgments}




\begin{thebibliography}{00}
\bibitem{fs83}  L. L. Frankfurt and M. I. Strikman, 
                   Nucl. Phys. A {\bf 405}, 557 (1983).
\bibitem{hjm89} P. Hoodbhoy, R. L. Jaffe, and A. Manohar,
                   Nucl. Phys. B {\bf 312}, 571 (1989);
                R. L. Jaffe and A. Manohar, Nucl. Phys. {\bf B321}, 343 (1989).
\bibitem{b1-sum} 
      F. E. Close and S. Kumano, Phys. Rev. D  {\bf 42}, 2377 (1990).
      The sum rule is based on the parton model explained
      in R. P. Feynman,
       {\it Photon-Hadron Interactions} (Westview press, Boulder, 1998). 
\bibitem{kh91} H. Khan and P. Hoodbhoy,
               Phys. Rev. C {\bf 44}, 1219 (1991).
\bibitem{b1-shadowing}
       N. N. Nikolaev and W. Sch\"afer,
                 Phys. Lett. B {\bf 398}, 245 (1997);
       Erratum, {\it ibid.}, B {\bf 407}, 453 (1997); 
       K. Bora and R. L. Jaffe, Phys. Rev. D {\bf 57}, 6906 (1998). 
\bibitem{epw-1997}       
       J. Edelmann, G. Piller, and W. Weise,
                 Z. Phys. A {\bf 357}, 129 (1997).
\bibitem{miller-b1} 
      G. A. Miller,  Phys. Rev. C {\bf 89}, 045203 (2014).     
\bibitem{pd-drell-yan}
      S. Hino and S. Kumano, Phys. Rev. D {\bf 59}, 094026 (1999);
                                          {\bf 60}, 054018 (1999);
      S. Kumano and M. Miyama, Phys. Lett. B {\bf 479}, 149 (2000).
\bibitem{ks-2016} S. Kumano and Qin-Tao Song, Phys. Rev. D {\bf 94}, 054022 (2016). 
\bibitem{Fermilab-dy} 
      Fermilab E1039 experiment, Letter of Intent Report No. P1039 (2013), 
      https://www.fnal.gov/directorate\\ 
      /program\_planning/June2013PACPublic/P-1039\_LOI \\ \_polarized\_DY.pdf;
     X. Jiang, D. Keller, A. Klein, and K. Nakano (private communication).
     For the on-going Fermilab E-906/SeaQuest experiment, see
           http:// \\ www.phy.anl.gov/mep/drell-yan/.
\bibitem{rho-production}
      A. Bacchetta and P. J. Mulders, Phys. Rev. D {\bf 62}, 114004 (2000). 
\bibitem{spin-1-frag}
      A. Sch\"afer, L. Szymanowski, and O. V. Teryaev,
                  Phys. Lett. B {\bf 464}, 94 (1999);
      K.-B. Chen, W.-H. Yang, S.-Y Wei, and Z.-T Liang,       
                  Phys. Rev. D {\bf 94}, 034003 (2016).
\bibitem{spin-1-gpd} 
      E. R. Berger, F. Cano, M. Diehl, and B. Pire, 
                  Phys. Rev. Lett. {\bf 87}, 142302 (2001);
      A. Kirchner and D. Mueller,
                  Eur. Phys. J. C {\bf 32}, 347 (2003); 
      M. Diehl, Phys. Rep. {\bf 388}, 41 (2003);
      F. Cano and B. Pire, Eur. Phys. J. A {\bf 19}, 423 (2004);
      A. V. Belitsky and A. V. Radyushkin, 
                Phys. Rep. {\bf 418}, 1 (2005). 
\bibitem{mass-corr} W. Detmold, Phys. Lett. B {\bf 632}, 261 (2006).
\bibitem{dmitrasinovic-96}
      V. Dmitrasinovic, Phys. Rev. D {\bf 54}, 1237 (1996).
\bibitem{lattice} C. Best {\it et al.}, 
                   Phys. Rev. D {\bf 56}, 2743 (1997).
\bibitem{angular-spin-1}	
     S. K. Taneja, K. Kathuria, S. Liuti, and G. R. Goldstein,
                Phys. Rev. D {\bf 86}, 036008 (2012).
\bibitem{tagged-spin-1}
     W. Cosyn, M. Sargsian, and C. Weiss, Proc. Sci. DIS2016 (2016) 210;
     to be published.
\bibitem{trans-g}
     R. L. Jaffe and A. Manohar, Phys. Lett. B {\bf 223}, 218 (1989);
     J. P. Ma, C. Wang, and G. P. Zhang, arXiv:1306.6693 [hep-ph];
     W. Detmold {\it et al.}, Letter of Intent Report No. LOI12-14-001 to
        Jefferson Lab PAC 42 (2014),
        https://www.jlab.org/exp\_prog/proposals \\ /14prop.html;
     W. Detmold and P. E. Shanahan, Phys. Rev. D {\bf 94}, 014507 (2016).
\bibitem{hermes05}
      A. Airapetian {\it et al.} (HERMES Collaboration), 
                   Phys. Rev. Lett. {\bf 95}, 242001 (2005).
\bibitem{tensor-pdfs}
      S. Kumano, Phys. Rev. D {\bf 82}, 017501 (2010).
\bibitem{basic-conv}
  G. L. Li, K. F. Liu, and G. E. Brown,
                        Phys. Lett. B {\bf 213}, 531 (1988);
  S. Kumano and F. E. Close, Phys. Rev. C {\bf 41}, 1855 (1990);
  M. M. Sargsian, S. Simula, and M. I. Strikman, 
                  Phys. Rev. C {\bf 66}, 024001 (2002);
  C. Ciofi degli Atti, L. L. Frankfurt, L. P. Kaptari, and M. I. Strikman, 
                  Phys. Rev. C {\bf 76}, 055206 (2007);
  M. Hirai, S. Kumano, K. Saito and T. Watanabe,
                  Phys. Rev. C {\bf 83}, 035202 (2011).
\bibitem{ek-2003} M. Ericson and S. Kumano,
                       Phys. Rev. C {\bf 67}, 022201 (2003).
\bibitem{nuclear-sfs} For example, see 
    D. F. Geesaman, K. Saito, and A. W. Thomas,
                        Ann. Rev. Nucl. Part. Sci. {\bf 45}, 337 (1995);
    R. G. Roberts,   {\it The Structure of the Nucleon}
                 (Cambridge University Press, 1993), pp. 8-12 \& 144-153. 
\bibitem{Frankfurt:1981mk}
    L. L. Frankfurt and M. I. Strikman, Phys. Rep. {\bf 76},  215  (1981).
\bibitem{Keister:1991sb}
    B. D. Keister and W. Polyzou, Adv. Nucl. Phys. {\bf 20}, 225 (1991).
\bibitem{cms-vna}     
    W. Cosyn and M. Sargsian,
           Phys. Rev. C {\bf 84}, 014601 (2011);   
    W. Cosyn, W. Melnitchouk, and M. Sargsian
           Phys. Rev. C {\bf 89}, 014612 (2014).       
\bibitem{Jlab-b1} Proposal to Jefferson Lab PAC-38 (PR12-11-110), 
                         J.-P. Chen {\it et al.} (2011),
                  https://www.jlab.org/exp\_prog \\ /proposals/11prop.html.
\bibitem{azz} W. Cosyn and M. Sargsian, J. Phys. Conf. Ser. {\bf 543}, 012006 (2014);
     M. M. Sargsian and M. I. Strikman, J. Phys. Conf. Ser. {\bf 543}, 012009 (2014);
                               E. Long, J. Phys. Conf. Ser. {\bf 543}, 012010 (2014);
     T. Badman {\it et al.}, Letter of Intent Report No. LOI12-14-002
        to Jefferson Lab PAC42, https://www.jlab.org/exp\_prog/proposals \\ /14prop.html.
\bibitem{eic} 
     W Cosyn {\it et al.}, J. Phys. Conf. Ser. {\bf 543}, 012007 (2014);
     N. Kalantarians, J. Phys. Conf. Ser. {\bf 543}, 012008 (2014);
          D. Boer {\it et al.}, arXiv:1108.1713 (unpublished);
          A. Accardi {\it et al.}, Eur. Phys. J. A {\bf 52}, 268 (2016);
          J. L. Abelleira Fernandez {\it et al.},
                J. Phys. G: Nucl. Part. Phys. 39 (2012) 075001.
\bibitem{kk08} T.-Y. Kimura and S. Kumano, Phys. Rev. D {\bf 78}, 117505 (2008).
\bibitem{sk14} S. Kumano, J. Phys.: Conf. Series {\bf 543}, 012001 (2014).
\bibitem{edmond} A. R. Edmond, {\it Angular Momentum in Quantum Mechanics}
                 (Princeton University Press, Princeton, 1974).
\bibitem{delta-T-notation} The overall factor 1/2 is introduced in $b_1$
      as it is used in $F_1$ and $g_1$ in expressing them by the unpolarized
      or longitudinally polarized PDFs.
\bibitem{flavor3} S. Kumano, Phys. Rep. {\bf 303}, 183 (1998);
                  G. T. Garvey and J.-C. Peng,
                       Prog. Part. Nucl. Phys. {\bf 47}, 203 (2001);
	   J.-C. Peng and J.-W. Qiu, Prog. Part. Nucl. Phys. {\bf 76}, 43 (2014).
\bibitem{hkn07} 
    M. Hirai, S. Kumano, and M. Miyama, Phys. Rev. D {\bf 64}, 034003 (2001);
    M. Hirai, S. Kumano, and T.-H. Nagai, Phys. Rev. C {\bf 70}, 044905 (2004);
                   {\bf 76}, 065207 (2007).
\bibitem{hermes-r} K. Ackerstaff {\it et al.} (HERMES Collaboration),
       Phys. Lett. B {\bf 475}, 386 (2000);
       {\bf 567}, 339(E) (2003).
\bibitem{Leader:2001gr} E. Leader, {\it Spin in Particle Physics} 
                         (Cambridge University Press, Cambridge, 2005).
\bibitem{relative-momentum} See Eq.\,(2.21) of Ref.\,\cite{Frankfurt:1981mk}.
\bibitem{cd-bonn-2001}
   R. Machleidt, Phys. Rev. C {\bf 63}, 024001 (2001).
\bibitem{MSTW2008} A. D. Martin, W. J. Stirling, R. S. Thorne,
                 and G. Watt,  Eur. Phys. J. C {\bf 63}, 189 (2009).
\bibitem{r1998} K. Abe {\it et al.} (E143 Collaboration), 
                   Phys. Lett. B {\bf 452}, 194 (1999).
\bibitem{d-energy}
   G. Audia, A. H. Wapstrab, and C. Thibaulta, 
                 Nucl. Phys. {\bf A729}, 337 (2003).
\bibitem{Bodek-etal-1979}
      A. Bodek {\it et al.}, Phys. Rev. D {\bf 20}, 1471 (1979).
\bibitem{duality} W. Melnitchouk, R. Ent, and C. Keppel,
                Phys. Rep. {\bf 406}, 127 (2005).
\bibitem{j-parc}
      S. Kumano,  Int. J. Mod. Phys.: Conf. Series, {\bf 40}, 1660009 (2016).
      see also contributions to Workshop on Hadron physics with high-momentum 
      hadron beams at J-PARC in 2013 and 2015,  
          http://www-conf.kek.jp/past/hadron1 \\ /j-parc-hm-2013/,
          http://research.kek.jp/group \\ /hadron10/j-parc-hm-2015/.
\bibitem{ilc} For information on the ILC project, see
                http://www \\ .linearcollider.org.
    On the TESLA-N, 
    see http://tesla \\ .desy.de/new\_pages/TDR\_CD/start.html.
\end{thebibliography}
\end{document}